\documentclass[sigplan,twocolumn]{acmart}

\usepackage{algorithm}
\usepackage{algorithmic}
\usepackage{pifont}

\renewcommand\footnotetextcopyrightpermission[1]{}
\AtBeginDocument{%
  }

\acmConference[preprint]{}{2025}{}

\begin{document}

\newcommand{\sysname}{OOCO}

\title{\sysname: Latency-disaggregated Architecture for Online-Offline Co-locate LLM Serving}


\author{Siyu Wu}
\affiliation{
  \institution{Beihang University}
  \city{Beijing}
  \country{China}
}
\email{wusiyu@buaa.edu.cn}

\author{Zihan Tang}
\affiliation{
  \institution{Tsinghua University}
  \city{Beijing}
  \country{China}
}
\email{tzh21@mails.tsinghua.edu.cn}

\author{Yuting Zeng}
\affiliation{
  \institution{University of Science and Technology of China}
  \city{Beijing}
  \country{China}
}
\email{yuting_zeng@mail.ustc.edu.cn}

\author{Hui Chen}
\affiliation{
  \institution{Tsinghua University}
  \city{Beijing}
  \country{China}
}
\email{jichenhui2012@gmail.com}

\author{Guiguang Ding}
\affiliation{
  \institution{Tsinghua University}
  \city{Beijing}
  \country{China}
}
\email{dinggg@tsinghua.edu.cn}

\author{Tongxuan Liu}
\authornote{Corresponding author.}
\affiliation{
  \institution{JD Company}
  \city{Beijing}
  \country{China}
}
\email{liutongxuan1@jd.com}

\author{Ke Zhang}
\authornotemark[1]
\affiliation{
  \institution{JD Company}
  \city{Beijing}
  \country{China}
}
\email{zhangke323@jd.com}

\author{Hailong Yang}
\authornotemark[1]
\affiliation{
  \institution{Beihang University}
  \city{Beijing}
  \country{China}
}
\email{hailong.yang@buaa.edu.cn}


\renewcommand{\shortauthors}{Siyu Wu et al.}

\begin{abstract}
Large Language Models (LLMs) are increasingly deployed in both latency-sensitive online services and cost-sensitive offline workloads. Co-locating these workloads on shared serving instances can improve resource utilization, but directly applying this approach to Prefill/Decode (P/D) disaggregated systems introduces severe load imbalance, as fluctuating request mixes alter the intrinsic P/D ratio. Existing dynamic adjustment techniques cannot keep up with the bursty traffic patterns of online services.

We propose a latency-constraint disaggregated architecture, which separates cluster resources into latency-strict and latency-relaxed pools based on task latency requirements. This design enables flexible placement of offline decode tasks, mitigating P/D imbalance while preserving online performance. To fully exploit this flexibility, we propose (1) a bottleneck-based scheduler guided by a Roofline-based performance model for performance bottleneck based scheduling, and (2) a fast preemption mechanism that strictly enforces Service Level Objectives (SLOs) for online requests.

Experiments on real-world traces show that compared to existing offline system approaches, our method improves offline throughput by up to 3×, while maintaining online request SLOs.

\end{abstract}

\maketitle

\section{Introduction}

The explosive growth of Large Language Models (LLMs) is profoundly transforming human-computer interaction and information processing. Recently, they have been widely applied in domains such as chatbot~\cite{ouyang2022training, touvron2023llama, claude, openai2023gpt4, guo2025deepseek}, code completion~\cite{li2023starcoder, copilot, hui2024qwen2}, translation~\cite{xu2024contrastive}, role-playing~\cite{shanahan2023role}, recommendation systems~\cite{deng2025onerec}, agent systems~\cite{wu2023tidybot}, intelligent data processing~\cite{chen2024data}, and dataset annotation~\cite{wang2024human}, gradually integrating into everyday life. This has led to continuously increasing and diversifying demands for LLM inference services. However, for LLM serving systems, although numerous optimizations in recent years have significantly improved efficiency, the inference cost of LLMs remains much higher than that of traditional services. As a result, improving efficiency or reducing cost has consistently been an important research direction for LLM serving system, especially as the application scope of LLMs continues to expand.

Although LLM services are diverse, they can generally be categorized into two types based on service form: online services and offline services.

\begin{figure}[t]
  \centering
  \includegraphics[width=1\linewidth]{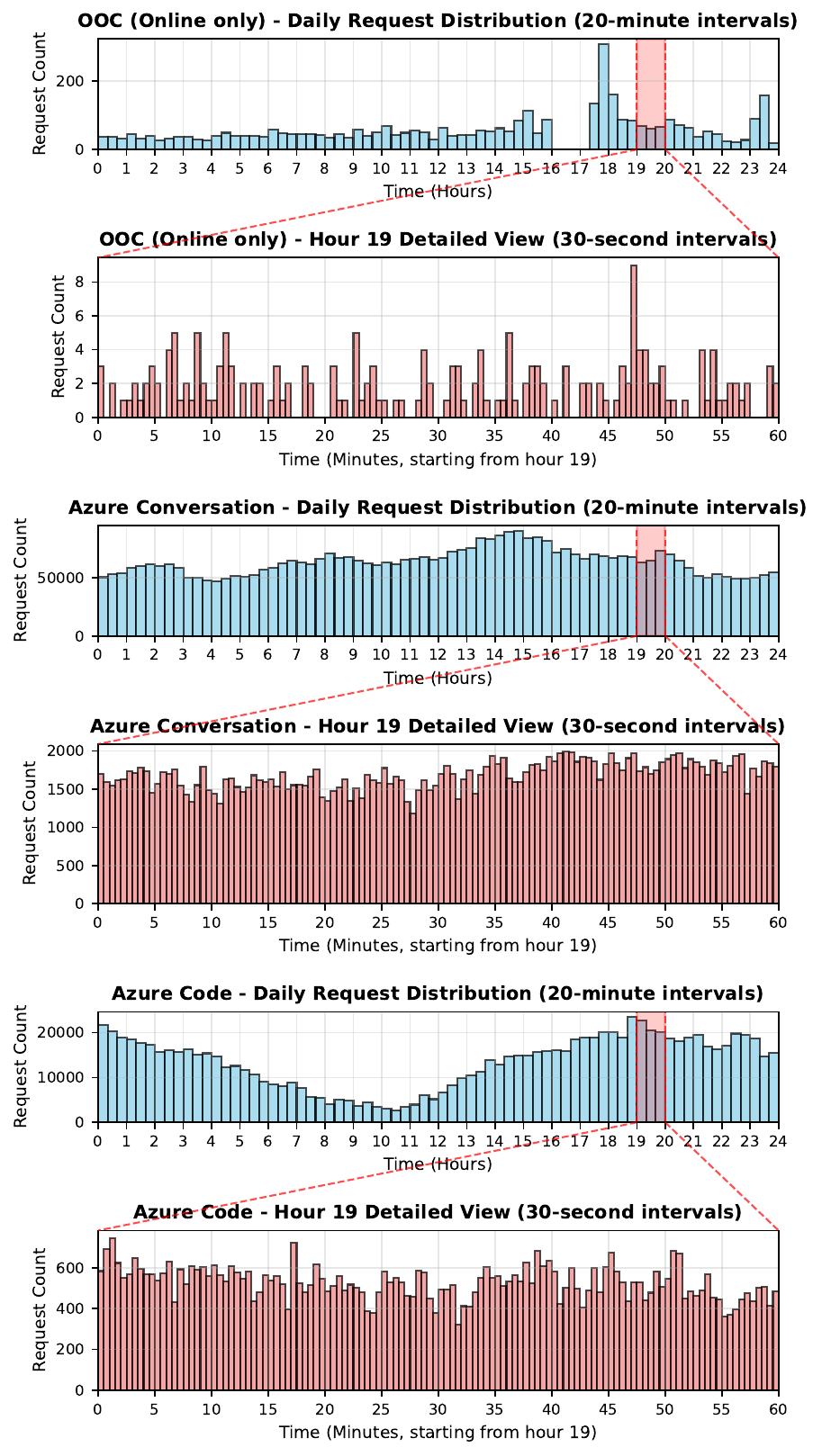}
  \caption{Visualization of request traffic variations patterns, showing tide-like fluctuations at hourly and daily scales, as well as bursty spikes at minute scales, across three datasets: the online service part of our company’s OOC trace, Azure LLM Inference Trace 2024 (Conversation), and Azure LLM Inference Trace 2024 (Code).}
  \label{fig:req_rate}
\end{figure}

\textbf{Online services}, such as chatbots, code completion, recommendation systems, virtual assistants, real-time translation, and interactive role-playing, are latency-sensitive workloads that require prompt responses upon receiving requests. These services often adopt streaming output to deliver each decoded token to the user in real time. As a result, they impose strict Service Level Objectives (SLOs) on both Time-To-First-Token (TTFT) and Time-Per-Output-Token (TPOT) to ensure user experience. Meanwhile, the request traffic of online services (i.e., request arrival rate) usually exhibits significant fluctuations, including tide-like variations at hourly or daily scales, as well as bursty spikes at minute scales, as shown in Figure\ref{fig:req_rate}. To meet SLOs under peak traffic, online service clusters often need to reserve enough resources for the worst-case demand, which causes substantial waste during off-peak periods. Although cluster auto-scaling can theoretically mitigate resource waste caused by tide-like variations, the slow cold-start of LLM service instances due to the large model weights loading and complex initialization~\cite{fu2024serverlessllm} makes it unsuitable for unpredictable bursty spikes. This creates risks of service failure, limiting the practical adoption of such techniques in industry.

\textbf{Offline services}, such as batch document processing, intelligent data analytics, and data annotation, are typically batch workloads with weaker real-time requirements. They do not impose constraints on TTFT or TPOT for individual requests, which grants higher flexibility for their serving. On the other hand, since the data volume can be very large, these services are usually highly cost-sensitive.

With the continuous advancement of model capabilities, a sufficiently powerful general-purpose model can now be adapted to a wide range of tasks, even simultaneously covering both online and offline services, through methods such as prompt engineering~\cite{giray2023prompt}, retrieval-augmented generation (RAG)~\cite{lewis2020retrieval}, and function calling, without requiring modifications to the LLM model itself. This makes it possible to run online and offline requests on the same LLM serving instance (the minimal unit that holds the complete model parameters and can independently perform model Prefill or Decode).

In such a co-located serving system, offline requests can leverage the idle resources during off-peak periods of online workloads. Compared with deploying online and offline services separately, this approach can effectively improve resource utilization, alleviate resource idleness caused by online request troughs, and reduce the cost of offline requests.

Existing co-located LLM serving systems for online and offline workloads are all based on architectures without Prefill / Decode (P/D) disaggregation. However, the P/D-disaggregated architecture, which decouples the two stages with very different computational characteristics, has demonstrated superior latency performance. As a result, it has appeared in the inference architecture reports of well-known companies such as MoonCake\cite{qin2024mooncake} and DeepSeek\cite{guo2025deepseek}, and is gradually becoming the mainstream approach in industry.

However, directly applying online-offline co-location to a P/D-disaggregated system introduces the problem of P/D load imbalance: a P/D-disaggregated system requires the load ratio between Prefill and Decode stages to match the resource ratio allocated to them. Otherwise, one stage can become a bottleneck, causing the other stage to stall or starve. Since online and offline requests usually have different P/D load ratios because of different service type, fluctuations in online workload shift the proportion of online versus offline requests, thereby changing the overall P/D load ratio and leading to imbalance. Moreover, because such overall P/D load ratio variation exhibits the same high fluctuation speed and magnitude as online traffic itself, it is difficult to resolve this issue with existing dynamic P/D adjustment techniques.

To address this problem, we rethink the design of P/D-disaggregated systems under online-offline co-location. The latency benefits of P/D-disaggregation essentially come from latency-constraint disaggregation: Decode is highly sensitive to per-step latency and cannot be blocked by long operations, which is why it must be decoupled from Prefill.

\begin{table}[h]
\centering
\footnotesize
\caption{Task placement under latency constraints. *Note that although online Prefill does not require sustained low latency environment like online Decode, its blockage can still lead to increased TTFT. We address this issue through a preemption mechanism.}
\renewcommand{\arraystretch}{1.2}
\setlength{\tabcolsep}{4pt}
\label{tab:tasks}
\begin{tabular}{lcccc}
\toprule
\textbf{Property} & \multicolumn{2}{c}{\textbf{Online}} & \multicolumn{2}{c}{\textbf{Offline}} \\
\cmidrule(lr){2-3} \cmidrule(lr){4-5}
 & Prefill & Decode & Prefill & Decode \\
\midrule
Intrinsic Latency   & \textcolor{red}{High} & \textcolor{green}{Low} & \textcolor{red}{High} & \textcolor{green}{Low} \\
Latency Sensitive   & \textcolor{red}{\ding{55}}* & \textcolor{green}{\ding{51}} & \textcolor{red}{\ding{55}} & \textcolor{red}{\ding{55}} \\
Placed on Latency Relaxed node        & \textcolor{green}{\ding{51}} & \textcolor{red}{\ding{55}} & \textcolor{green}{\ding{51}} & \textcolor{green}{\ding{51}} \\
Placed on Latency Strict node         & \textcolor{red}{\ding{55}} & \textcolor{green}{\ding{51}} & \textcolor{red}{\ding{55}} & \textcolor{green}{\ding{51}} \\
\bottomrule
\end{tabular}
\end{table}

Inspired by this, we propose a \textbf{latency-constraint disaggregated architecture} for the online-offline co-location scenario. In this design, cluster resources are divided into a latency-relaxed pool and a latency-strict pool, and all tasks are reassigned to the pools based on their intrinsic latency characteristics and latency requirements, as illustrated in Table~\ref{tab:tasks}. In this architecture the Decode phase of offline requests can be executed in either resource pool, which grants us the ability to flexibly adjust the load ratio between the two pools and thereby maximize overall cluster resource utilization.

However, this architecture also introduces the following two challenges:

\textit{1) More complex scheduling space}: Since the Decode phase of offline requests can be executed on either type of instance, a non-trivial question arises: how to leverage this flexibility to design intelligent scheduling strategies that further optimize efficiency. Compared with simple heuristic approaches, proposing an optimal scheduling policy is a challenging task.

\textit{2) Challenges to strict SLO guarantees}: Offline requests inevitably consume resources. For example, their Prefill stage may block newly arriving online requests, or their Decode stage on latency-strict nodes may slow down overall response time. Both cases can negatively impact the SLO of online requests. Ensuring that the SLO satisfaction rate remains equivalent to that of a purely online P/D-disaggregated system requires careful and precise consideration.

To address the first challenge, we propose a performance bottleneck based scheduler to achieve fine-grained control over request scheduling. We build an LLM inference performance model based on the Roofline Model~\cite{williams2009roofline}. This model predicts the latency as well as compute and memory utilization of both Prefill and Decode stages. Since Decode phase on latency-strict instances usually account for a large share of execution time and are highly performance-sensitive, we target a balance between compute and memory on latency-strict instances as the optimization goal. By analyzing performance bottlenecks with this model, we can select more suitable offline requests to merge into Decode batches.

To address the second challenge, we propose a fast preemption mechanism to guarantee online SLOs to strictly ensure that online SLOs are not compromised. We introduce a preemption mechanism of online requests over offline requests. For offline Prefill tasks running on latency-relaxed nodes, we propose a layer level interruption technique, which enables preemption within acceptable latency bounds without incurring additional future LLM model supporting costs. For Decode tasks on latency-strict nodes, we leverage the performance model to dynamically select Decode batch requests, ensuring that Decode latency consistently meets SLO constraints.

In summary, our work makes the following contributions:
\begin{itemize}
\item We propose a latency-constraint disaggregated architecture for the online-offline co-location scenario, which achieves SLO performance for online requests equivalent to that of a P/D-disaggregated system, while providing higher resource utilization.
\item We design a bottleneck-based scheduler, with a Roofline-based performance model for LLM inference, to achieve compute–memory balanced scheduling.
\item We based on a high-performance C++ LLM inference framework, xLLM, and implement \sysname{} on top of it for efficiency online-offline co-located serving.
\item We evaluate the system on datasets from real-world co-location scenarios, demonstrating that \sysname{} achieves higher offline throughput performance while maintaining the SLO of online requests.
\end{itemize}

\section{Background and Motivation}

In this section, we introduce the background of LLM inference systems and our analysis of performance bottlenecks in LLM inference.

\subsection{LLM Inference Systems}

Unlike traditional DNNs that perform a single forward pass, LLMs generate outputs in an auto-regressive manner, requiring repeated inference for each output token~\cite{radford2019language}. 

To avoid re-computing the entire input at every step, modern systems cache intermediate attention states, known as the KV cache. This design naturally divides inference into two stages: Prefill, which processes the full input sequence to build the KV cache (keep the K and V tensor in echo attention blocks) and generate the first token, and Decode, which only the most recent token is fed into the model for inference, reuses and expands the KV cache to generate a new token once per Decode execution.

Since LLM requests have varying lengths, they cannot be batched in the same way as traditional DNNs, as padding would lead to substantial resource waste. To address this, ORCA~\cite{yu2022orca} introduced the concept of continuous batching, which is now supported by most mainstream high-performance LLM inference frameworks. In practice, the continuous batching is typically achieved by combining a non-contiguous KV cache memory pool~\cite{kwon2023efficient} with customized attention kernels, enabling LLM serving systems to batch requests of different lengths together. Requests can be dynamically added to or removed from the batch at each batch Decode execution (or "step"), providing iteration-level scheduling granularity (i.e., per Prefill or Decode execution).

In LLM service deployment, the Service Level Objectives (SLOs) are critical. Online applications such as chatbots and code completion require strict latency guarantees, commonly measured by Time-To-First-Token (TTFT) and Time-Per-Output-Token (TPOT), mainly corresponding to the inference time of Prefill and each Decode execution. Offline tasks such as batch analytics are more latency-tolerant but highly cost-sensitive, creating different design pressures for inference systems.

One important architectural advancement is the Prefill / Decode (P/D) disaggregated design~\cite{zhong2024distserve}. Traditional inference systems process Prefill and Decode within the same instance, which leads to inefficiency, as the interactions between Prefill and Decode make it harder to meet SLOs. P/D-disaggregated architecture decouples these two stages into separate instances or resource pools, allowing more fine-grained scheduling and tailored resource provisioning. Decode steps, which are highly latency-sensitive, can be isolated from the relatively high latency Prefill steps, avoiding head-of-line blocking and ensuring lower per-step latency. Meanwhile, Prefill operations can be scheduled independently to exploit batch-level parallelism. This separation has been shown to improve system efficiency and SLO adherence, and is increasingly adopted in both academic and industrial systems.

\subsection{Online Workload Fluctuations and Offline Co-location}
Online LLM services, such as chatbots or code assistants, face highly dynamic traffic patterns. Request arrival rates fluctuate at multiple time scales: long-term tide-like variations (hourly or daily) and short-term bursty spikes (seconds to minutes), as shown previously in Figure~\ref{fig:req_rate}. To satisfy strict SLOs under peak load, clusters are often provisioned with sufficient resources for the worst case. This leads to significant resource idleness during off-peak periods.

In contrast, offline tasks such as batch data processing or annotation have much looser latency requirements. They can flexibly utilize available capacity but are extremely cost-sensitive due to large data volumes. This complementary nature makes online-offline co-location attractive: idle resources left unused by fluctuating online demand can be reclaimed to process offline workloads, improving cluster utilization and reducing inference cost.

\subsection{LLM Inference Performance Character}

\begin{figure}[t]
  \centering
  \includegraphics[width=0.8\linewidth]{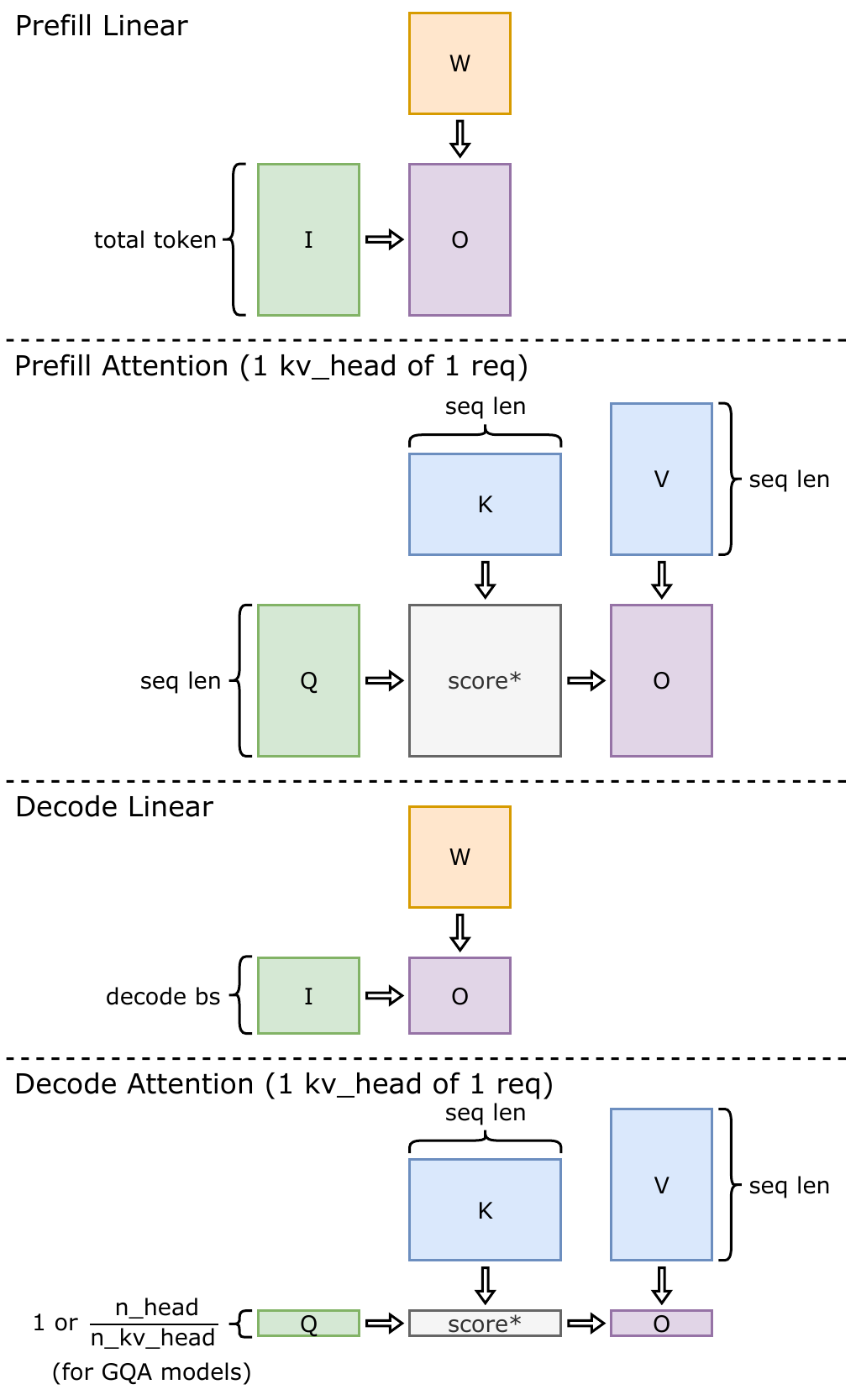}
  \caption{The computation patterns of the main time-consuming operators in LLMs. *In the figure, non-matrix multiplication operations such as Softmax in the Attention operator are omitted. Additionally, for the commonly used Flash Attention~\cite{dao2022flashattention, dao2023flashattention2, hong2023flashdecoding++} operator, the intermediate score matrix is passed through on-chip buffers or cache, without generating memory accesses to the GPU / NPU memory.}
  \label{fig:llmop}
\end{figure}

In an online-offline co-located system, offline requests must be precisely controlled so that system resources are fully utilized without compromising the SLOs of online requests. Achieving this requires a clear understanding of the performance characteristics of LLM inference, including both latency and the balance between computation and memory access. LLM inference involves substantial compute and memory operations, and its Prefill and Decode phases exhibit fundamentally different computational patterns, as shown in Figure~\ref{fig:llmop}. Their latency is further influenced by factors such as batch size and request length.

\begin{figure}[t]
  \centering
  \includegraphics[width=\linewidth]{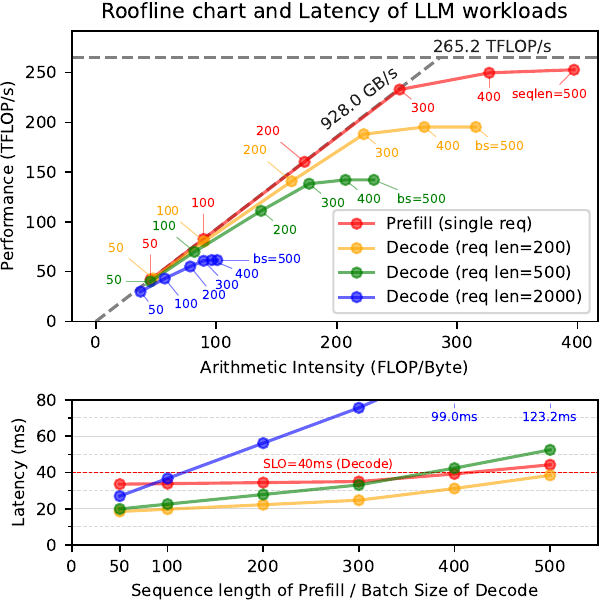}
  \caption{Roofline analysis with corresponding latency of LLM inference (Qwen2.5 7B on Ascend 910c NPU). Each point denotes a Prefill or Decode execution under a given batch size and request length.}
  \label{fig:fig_llm_perf}
\end{figure}

In Figure~\ref{fig:fig_llm_perf}, we present the roofline analysis and corresponding latency of LLM inference. Each point represents a Prefill or Decode execution (a single model forward) under a specific batch size and request length. Results are obtained on Qwen2.5 7B~\cite{team2024qwen2} (at bf16) running on the Ascend 910c NPU~\cite{zuo2025serving} using our performance model, which accurately aligns with actual execution in xLLM, including runtime overhead. The peak FLOPs and memory bandwidth in the roofline chart are achievable values measured through \sysname{}’s profiling data.

The Prefill phase resembles the standard forward pass of a decoder-only Transformer model~\cite{vaswani2017attention}. Every token in the input sequence participates in the computation of linear layers and attention modules. Memory accesses are dominated by model weights and intermediate activations, with the K and V tensors preserved as the KV cache. Prefill thus tends to achieve high compute intensity, and once the sequence length reaches around 250 tokens on 910c, execution becomes compute-saturated, classifying Prefill as compute-intensive.

The Decode phase, however, differs due to the KV cache mechanism. Here, each request in the batch processes only a single token per step. The effective input size for linear layers and the attention Q tensor is therefore determined by the batch size, rather than the total sequence length across requests. Since each request maintains its KV cache in GPU memory, the Decode batch size is constrained by memory capacity. Memory accesses include not only model weights and activations but also the entire KV cache of all requests in the batch, making Decode generally memory-intensive. However, modern compression techniques such as MQA~\cite{shazeer2019fast}, GQA~\cite{ainslie2023gqa}, and MLA~\cite{liu2024deepseek} significantly reduce KV cache size, allowing Decode batch sizes to scale into the hundreds even for moderate request lengths. Because KV cache accesses occur only in the attention operators, while linear layers remain compute-intensive (with Decode batch size = N comparable to Prefill sequence length = N), the linear layers can again become compute-bound, limiting further achieved FLOPs/s improvements.

From a latency perspective, Prefill with sequence length N and Decode with batch size N often exhibit similar latency when processing short requests (with Prefill slightly slower in xLLM due to runtime overhead). For longer requests, however, the growing KV cache size leads to significantly higher Decode latency. This illustrates that LLM inference latency cannot be determined solely by batch size or total sequence length, but must consider their combined effects along with KV cache overhead.

\section{Methodology}

\begin{figure*}[t]
  \centering
  \includegraphics[width=0.8\linewidth]{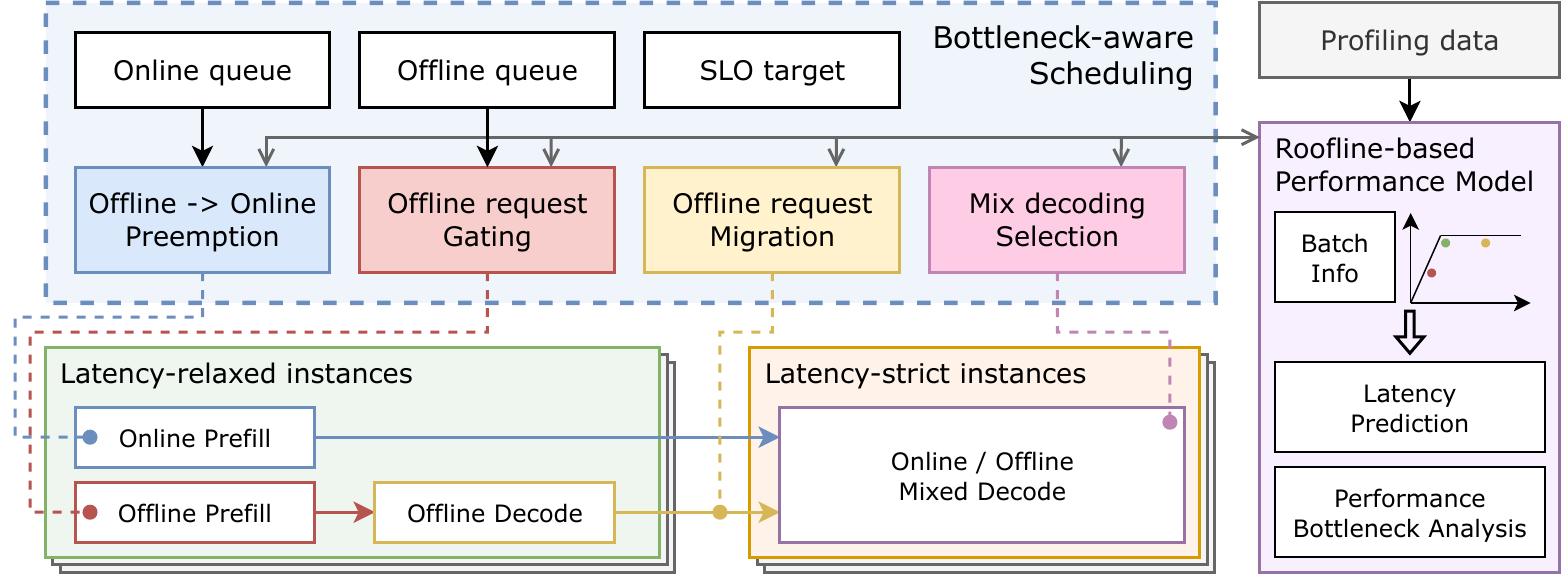}
  \caption{Overview of \sysname{}.}
  \label{fig:fig_main}
\end{figure*}


\subsection{Overall Architecture}
Figure~\ref{fig:fig_main} illustrates the overview of our proposed online-offline co-located LLM serving system: \sysname{} (xLLM Online-Offline Co-locate). The design follows the principle of latency-constraint disaggregation, partitioning cluster resources into latency-relaxed instances and latency-strict instances. Incoming online and offline requests scheduled onto these instances through bottleneck-aware scheduling leveraging our Roofline-based performance model. We will discuss these parts in detail below.

\subsection{Instance Type}
Each instance represents the minimal unit that can execute model forward independently, occupying one or more GPUs (or NPUs) depending on the model parallelism size. We divided instances into two types based on the latency constraints of online requests' TPOT SLO. 

\subsubsection{Latency-relaxed Instances}
The Latency-relaxed instances are those where the execution time of a single iteration is not restricted. In other words, Prefill and Decode steps with arbitrary latency can be executed on these instances. However, online Decode cannot be placed here, since long per-iteration latency would prevent TPOT SLO from being satisfied. Thus, latency-relaxed instances are used for online/offline Prefill, and may also host offline Decode, which has no TPOT constraints.

\subsubsection{Latency-strict Instances}
Latency-strict instances, in contrast, enforce restrictions on the execution time of each iteration. Only Decode operations can be executed here, and their latency must stay below the TPOT SLO to ensure the responsiveness of online requests Decode. During off-peak periods of online requests, offline requests Decode can also be mixed-in on these instances to maximize resource utilization.

\subsection{Performance Model}

To enable efficient and SLO-compliant scheduling of LLM execution, we require the ability to predict the latency of a given request batch or to identify its performance bottleneck. This necessitates building performance models for the Prefill and Decode stages of LLM inference, capable of estimating the latency, computational workload, and memory traffic of any given batch.

Since the structure of common LLMs is relatively fixed, we construct an operator-level behavioral simulator for typical model architectures. This simulator models the computation and memory traffic of individual operators, and employs the roofline model to predict execution latency.

\subsubsection{Operator Modeling}

The majority of inference time is spent in linear layers (GEMM operators), attention operators, and communication operations. We model the GEMM and attention operators as shown in Table 2, with symbol definitions in Table 1. For computation, we use theoretical FLOP counts. For memory traffic, inspired by prior performance-analysis work such as PRoof~\cite{wu2024proof}, we assume operators can effectively utilize on-chip cache or buffers, and thus consider the total size of required input/output tensors as their memory access amount.

For the attention operator, since mainstream LLM inference frameworks (including xLLM) employ Flash Attention or similar fused kernels, intermediate attention-score tensors are stored in on-chip SRAM using tiling, avoiding excessive memory traffic for long sequences. We therefore model the fused attention kernel as a single operator, which better captures its actual runtime behavior.

\begin{table}
  \centering
  \footnotesize
  \caption{Symbol definitions.}
  \label{tab:symbol}
  \begin{tabular}{ll}
    \toprule
    Symbol & Define \\
    \midrule
    $d$ & Byte per value (e.g. 2 for float16) \\
    $N$ & Input size for GEMM (e.g. total token number for Linear) \\
    $D_{in}$ & Input dim for GEMM \\
    $D_{out}$ & Output dim for GEMM \\
    $D_h$ & Hidden dim for attention \\
    $S_q$ & Sequence length for attention Q tensor \\
    $S_{kv}$ & Sequence length for attention K or V tensor \\
    $H_q$ & Head number for Q \\
    $H_{kv}$ & Head number for K or V \\
  \bottomrule
\end{tabular}
\end{table}

\begin{table}
  \centering
  \footnotesize
  \caption{Operator Modeling.}
  \label{tab:op_model}
  \begin{tabular}{lll}
    \toprule
    Operator & Compute (FLOPs) & Memory (GB) \\
    \midrule

    GEMM & $2 N D_{in} D_{out}$ & $ d \cdot (N D_{in} + D_{in} D_{out} + N D_{out})$ \\
    Attention & $4 D_h S_q S_{kv}$ & $ 2 d \cdot (S_q D_h + S_{kv} D_h \cdot \frac{H_q}{H_{kv}}) $\\
  \bottomrule
\end{tabular}
\end{table}

\subsubsection{Roofline-based Latency Modeling}

\begin{table}
  \centering
  \footnotesize
  \caption{Parameters for latency modeling.}
  \label{tab:latency_p}
  \begin{tabular}{ll}
    \toprule
    Symbol & Define \\
    \midrule
    $F_g$ & Achievable FLOPs/s for GEMM operators\\
    $F_{ap}$ & Achievable FLOPs/s for Prefill Attention operators\\
    $F_{ad}$ & Achievable FLOPs/s for Decode Attention operators\\
    $M_g$ & Achievable memory bandwidth for GEMM operators\\
    $M_a$ & Achievable memory bandwidth for Attention operators\\
    $O_p$ & Static runtime overhead latency for Prefill \\
    $O_d$ & Static runtime overhead latency for Decode \\
    $B_c$ & Effective bandwidth for communication operator \\
  \bottomrule
\end{tabular}
\end{table}

Comparing with heuristic or ML-based approaches, our roofline-based simulation method fully leverages known model structures and prior knowledge, requires less runtime information and offers greater robustness. However, latency estimation still requires some environment-dependent parameters as shown in Table 3, which could be obtained through a small amount of profiling data.

$F_*$ and $M_*$ denote the achievable FLOPs/s and achievable memory bandwidth for each operator type, which are typically lower than the hardware’s theoretical peak. The Flops of the Attention operator are further subdivided into Prefill and Decode scenarios, as their distinct implementation approaches result in differing efficiencies in computational resource utilization. Incorporating these values into the roofline model, we derive the operator latency from its computation and memory traffic, as expressed in Equation \ref{eq:op_latency}.

\begin{equation}
\begin{split}
  Op\ Latency = max(\frac{Op\ FLOPs}{F_a}, \frac{Op\ Memory}{M_a})
\end{split}
\label{eq:op_latency}
\end{equation}

Moreover, $O_p$ and $O_d$ denote the static runtime overhead latency of a Prefill and a Decode iteration, respectively. This includes CPU-side logic, kernel launch time, and other fixed costs. Such overhead remains largely constant and does not vary with the batch size or sequence length of requests. 

$B_c$ denote effective interconnect bandwidth for communication operators. Their latency will be simply calculated by dividing the data volume by the effective bandwidth. The network delay is a fixed value and will be included in the static runtime overhead parameters.

The final latency prediction is obtained by summing the latency of all operators within the iteration and adding this overhead term. We validate the accuracy of our model on Qwen2.5 7B and 72B models, where the average absolute error is around 5\%, which is sufficient.

\subsubsection{Performance Bottleneck Analysis} The hardware capabilities of an accelerator card (e.g., GPU or NPU) can be divided into compute resources, memory bandwidth, and memory capacity. For maximum efficiency, we naturally aim to fully utilize all of these resources. However, in real inference workloads, due to the inference mode (Prefill or Decode), request count, and request length, the usage of different resources is often imbalanced. As observed in Figure~\ref{fig:fig_llm_perf}, such imbalance causes one or more resources (driven by different operator types in the inference process) to become bottlenecks, thereby limiting efficiency. In the following, we provide a detailed discussion and classification of these bottlenecks.

For Prefill, the situation is relatively straightforward. Unless the sequence length is very short (e.g., less than ~300 on the Ascend 910c NPU), Prefill typically falls into a compute bottleneck, dominated by GEMM and attention operators. Meanwhile, in most non-extreme text scenarios, memory capacity is sufficient for individual requests and does not become a limiting factor.

For Decode, the situation is more complex. The compute intensity varies significantly depending on the batch size and request length, and it is further constrained by memory capacity. As discussed in earlier sections, the linear (GEMM) operators in Decode may exhibit either memory-bound or compute-bound behavior depending on the batch size. In contrast, attention operators are typically memory-bound under the Decode computation mode. This leads to the following latency patterns:

\begin{itemize}
    \item The attention latency grows roughly in proportion to the total token count in the batch.
	\item When the Decode batch size is small (e.g., less than ~300 on the Ascend 910c NPU), GEMM latency remains relatively constant, making the workload memory-bound overall.
	\item When the Decode batch size is large, GEMM latency scales proportionally with batch size, leading to both compute and memory becoming bottlenecks simultaneously.
\end{itemize}

In addition, the total token count is restricted by memory capacity. On latency-strict nodes, the per-step Decode latency is also tightly constrained by the TPOT SLO, further complicating scheduling.

\subsection{Secluding Logic}
With the scheduling space provided by the latency-constraint disaggregated architecture and the performance bottleneck prediction from our performance model, we design scheduling policies that maximize resource utilization. We divide the request scheduling along the data path into four independent scheduling points: online request preemption, offline request gating, offline request migration, and mixed decoding selection. Their positions in the data path are illustrated by the arrows in Figure~\ref{fig:fig_main}.

\subsubsection{Online Request Preemption}
To guarantee the time-to-first-token (TTFT) of online requests, they must remain unaffected by offline requests, ensuring that their SLO performance matches that of a pure P/D-disaggregated system. To this end, we introduce a preemption mechanism where an online request Prefill can interrupt offline Prefill step. Specifically, when an online request arrives, if a latency-relaxed instance is currently running the Prefill or Decode of offline requests, it can be preempted immediately to serve the online Prefill.

To balance preemption speed and implementation cost (especially given that LLM inference frameworks constantly adapt to new model architectures), we choose the transformer-layer level as the granularity of interruption. This approach avoids modifying model-specific implementations while still achieving preemption within tens of milliseconds, which is acceptable compared to the seconds-level TTFT SLO. Moreover, this design facilitates future support for checkpoint-based recovery.

On the other hand, once an online request completes its Prefill stage, it must be quickly dispatched to a latency-strict instance for decoding. This requires sufficient free KV cache memory on the latency-strict nodes. If necessary, we evict some offline requests already running there. The choice of which offline requests to evict introduces another scheduling dimension:

\begin{itemize}

\item Evicting multiple short offline requests satisfies a given token-space requirement with relatively lower recomputation overhead, but reduces the available Decode batch size more significantly.
\item Evicting fewer long offline requests preserves a larger Decode batch size, but incurs higher recomputation costs.
\end{itemize}

Therefore, the selection of offline requests for eviction depends on the dominant performance bottleneck of the latency-strict node. If the bottleneck is computation, the scheduler prefers evicting longer requests to avoid excessively shrinking the Decode batch size.

\subsubsection{Offline Request Gating}
When a latency-relaxed node has no pending online requests to Prefill, it can either perform offline request Prefill or execute offline request Decode. If sufficient KV cache memory is available, the system may choose to Prefill new offline requests, thereby enlarging the Decode batch size for subsequent offline decoding. In certain cases, this improves efficiency. However, if these requests are later evicted during online preemption, additional recompute overhead will be incurred.

To balance these trade-offs, we design a cost model to guide the decision. Prefill for new offline requests is only performed if the effective latency reduction (achieved by increasing the future offline Decode batch size) exceeds the expected recompute overhead (caused by potential eviction). This ensures that offline gating decisions consistently contribute to overall system efficiency without jeopardizing online request SLO guarantees.

\subsubsection{Offline Request Migration}

Similar to online requests, offline requests can also transfer their KV cache from a latency-relaxed node to a latency-strict node for decoding. The difference is that for online requests, we want them to be transferred immediately after Prefill to start decoding in order to guarantee the SLO. For this purpose, we use a \textit{push model}, where latency-relaxed nodes actively search for suitable latency-strict nodes and push requests to them.

For offline requests, however, this is unnecessary. Latency-strict nodes will only process additional offline requests when the latency of the current decoding step still leaves room under the SLO, and they can proactively choose which offline requests to process. Therefore, we design a \textit{pull model}, where latency-strict nodes, upon determining it is appropriate, send a pull signal with their request length preference to a latency-relaxed node. The latency-relaxed node then select from the currently decoding offline requests the ones that best match the criteria and migrate them.

Specifically, the migration decision first reviews the mix decoding selection from the previous step (to be detailed in the next section). Migration is only triggered if, after all requests on the node have already been included in the decoding batch, the latency still remains within the TPOT SLO constraint with some margin. If this condition is met, the system evaluates the current performance bottleneck to determine the length preference for offline requests, with the goal of maximizing the utilization of compute and memory resources while staying within the SLO bound.

\begin{algorithm}
	\renewcommand{\algorithmicrequire}{\textbf{Input:}}
	\renewcommand{\algorithmicensure}{\textbf{Output:}}
	\caption{Offline Request Migration Decision}
	\label{alg:migration}
	\begin{algorithmic}[1]
		\REQUIRE $B$: Current decoding batch; 
		         $S$: TPOT SLO latency bound; 
		         $L(\cdot)$: Latency predictor; 
		         $bs(\cdot)$: Batch size function; 
		         $bs_{\text{sat}}$: compute-saturated batch size threshold; 
		         $U_{mem}(\cdot)$: Memory capacity utilization
		\ENSURE $Pref$: Length preference for migrated offline requests
		\STATE Review previous step's mix decoding selection
		\IF{$L(B) < S$ \AND all requests in node are included in $B$}
			\STATE Evaluate bottleneck type
			\IF{$bs(B) \ge bs_{\text{sat}}$} 
				\STATE $Pref \leftarrow \max \{ \ell \mid L(B \cup r) \leq S,\, U_{mem}(B \cup r) \leq 1 \}$ 
			\ELSE
				\IF{$\exists r:\ L(B \cup r) \leq S \ \AND\ bs(B \cup r) \ge bs_{\text{sat}}$}
					\STATE $Pref \leftarrow \max \{ \ell \mid L(B \cup r) \leq S \}$
				\ELSE
					\STATE $Pref \leftarrow \min \{ \ell \}$ \COMMENT{Choose shortest requests}
				\ENDIF
			\ENDIF
			\STATE Send pull request with $Pref$ to latency-relaxed node
			\STATE Latency-relaxed node selects ongoing offline requests most closed to $Pref$ and migrates them
		\ELSE
			\STATE $Pref \leftarrow \varnothing$ \COMMENT{No migration}
		\ENDIF
	\end{algorithmic}  
\end{algorithm}

As shown in Algorithm~\ref{alg:migration}, the procedure first checks whether the Decode batch has already reached compute saturation. If saturation is reached, it means that further increasing the batch size will not improve computational efficiency. In this case, the objective shifts to fully utilizing memory capacity, and the length preference is set to the longest request length that fits within both the SLO and memory capacity constraints.

If the batch has not yet reached compute saturation, the immediate goal becomes increasing the batch size to reach compute saturation. The system further checks whether adding a group of requests to the current batch has a chance to achieve compute saturation within the SLO constraint. If so, it calculates the maximum permissible request length under the SLO constraint and sets it as the length preference. Otherwise, the length preference is set to the shortest possible requests in order to maximize the batch size.

Although Offline Request Migration is only triggered when the current latency is below the SLO, LLM decoding is a dynamic process: request lengths continuously grow until completion and are inherently unpredictable. Therefore, before each decoding step, we must dynamically select which requests to include in the batch to ensure that the overall latency still meets the SLO requirement.

\subsubsection{Mix Decoding Selection}

\begin{algorithm}
	\renewcommand{\algorithmicrequire}{\textbf{Input:}}
	\renewcommand{\algorithmicensure}{\textbf{Output:}}
	\caption{Mix Decoding Selection}
	\label{alg:mix}
	\begin{algorithmic}[1]
		\REQUIRE $R_{on}$: Set of online requests; 
		         $R_{off}$: Set of candidate offline requests; 
		         $S$: TPOT SLO latency bound; 
		         $L(\cdot)$: Latency predictor
		\ENSURE $B$: Selected decoding batch
		\STATE Initialize $B \leftarrow R_{on}$  \COMMENT{All online requests must be included}
		\STATE Randomly iterate over $R_{off}$ up to $K$ times
		\FOR{each randomly selected $r \in R_{off}$}
			\IF{$L(B \cup \{r\}) \leq S$}
				\STATE $B \leftarrow B \cup \{r\}$
			\ELSE
				\STATE Discard $r$
			\ENDIF
		\ENDFOR
		\IF{$\exists \ r \in R_{off}$ not tested \AND $L(B) < S$}
			\STATE Sort remaining $R_{off}$ by ascending request length
			\STATE Apply binary search on sorted $R_{off}$ to find largest prefix $R' \subseteq R_{off}$ such that $L(B \cup R') \leq S$
			\STATE $B \leftarrow B \cup R'$
		\ENDIF
		\STATE \textbf{return} $B$
	\end{algorithmic}  
\end{algorithm}

All online requests are always included in the batch first. Then, offline requests are selected based on the SLO constraint. To avoid starvation, the scheduler iteratively and randomly selects offline requests until either a maximum number of iterations is reached or the SLO is exceeded. If the SLO is still not exceeded, the remaining offline requests are sorted in ascending order of length. This ensures that when only part of the offline requests can be included, the batch size is maximized. Finally, a binary search is applied to determine the exact subset of offline requests that can fit within the SLO bound, as illustrated in Algorithm~\ref{alg:mix}.

Cases where the latency of online requests already exceeds the SLO typically only occur when online requests are stalled, which is beyond the scope of this paper. However, in our framework, this can be configured either to ignore the SLO and still Decode all online requests (best-effort mode) or to sacrifice a portion of requests in order to preserve the SLO for the remaining ones.

\section{Implementation}

\sysname{} is implemented on top of the xLLM framework, a high-performance LLM inference engine written entirely in C++. The framework provides tier-1 support for NPU hardware and contains more than 100K lines of code. It incorporates key LLM optimization techniques, including continuous batching, paged attention, P/D-disaggregation, efficient operator implementations, and graph-fusion optimizations. On both NPUs and GPUs, xLLM achieves performance on par with, or even surpassing, mainstream frameworks such as vLLM and SGLang. xLLM has already been deployed across our company’s core retail businesses, supporting a wide range of applications such as intelligent customer service, risk control, supply chain optimization, and advertising recommendation.

\begin{figure}[t]
  \centering
  \includegraphics[width=\linewidth]{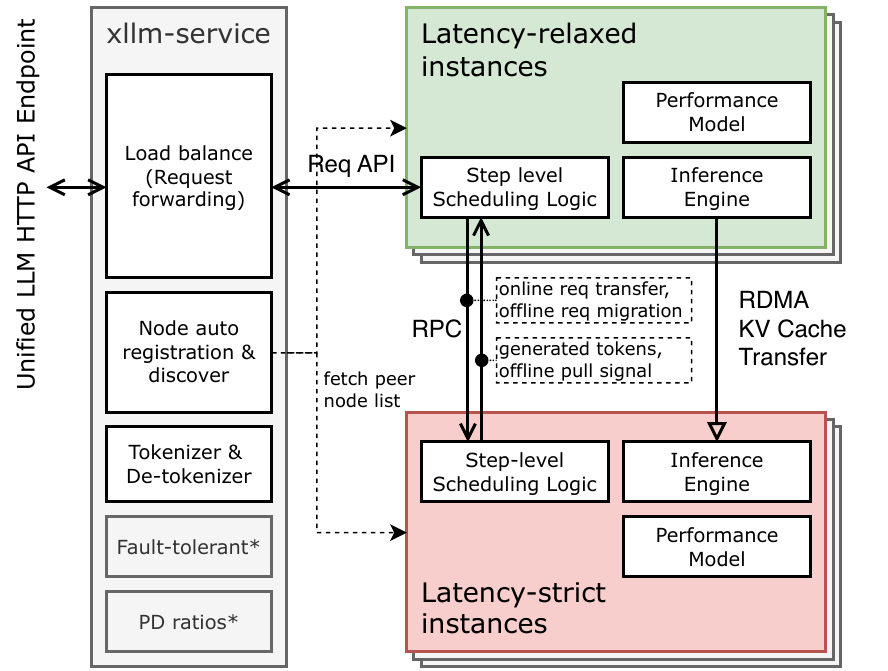}
  \caption{System architecture of \sysname{}. Components marked with * are supported by xLLM but are disabled in our implementation since they are orthogonal to this work (see Section~\ref{sec:discussion} for more details).}
  \label{fig:fig_impl}
\end{figure}

The framework consists of two major parts: the xllm-service layer and the xllm instance, as shown in Figure~\ref{fig:fig_impl}. An xllm instance is the minimal execution unit capable of performing one stage of LLM inference independently, and it encapsulates the main inference logic. The step-level scheduling logic and performance models of \sysname{} are implemented directly within xllm instances, enabling distributed scheduling and avoiding the scalability limitations of centralized schedulers. \sysname{} relies on the xllm-service for multi-instance request-level scheduling and load balancing, as well as for providing coordination between latency-relaxed and latency-strict instances. Migration and other control logics are then executed directly between xllm instances through RPC, with RDMA used for KV cache transmission. Importantly, the scheduling methods of \sysname{} are not tied to the specific features or architecture of xLLM, and can be readily ported to other mainstream LLM inference frameworks.

\section{Evaluation}

\subsection{Experimental Setup}

\subsubsection{Hardware}
Our experimental environment is based on nodes from the CloudMatrix 384 system~\cite{zuo2025serving}, each node contains 8x Ascend 910c NPUs, 4x Kunpeng 920 CPUs, and 3 TB of memory. The AI accelerator used is the Ascend 910c NPU, which adopts a dual-chip design. Each chip can be used independently and delivers theoretical performance metrics and practical LLM performance comparable to the NVIDIA A100 SXM. We use one latency-relaxed instance and one latency-strict instance for evaluations. For the 7B model we deploy each instance on a single chip, and for the 72B model we deploy each instance on 4 chips with tensor parallel.

\subsubsection{Model and Datasets}
We select Qwen2.5 7B and Qwen2.5 72B as our models, which provides sufficient capability for multi-business scenarios while maintaining relatively low deployment cost, and serves as a widely adopted base model in our production. The datasets involve both online and offline requests, with three configurations used in our experiments. Among them, OOC\footnote{we will soon open-source the de-identified dataset.} is derived from real traces of our company’s production workloads and, to the best of our knowledge, is the first dataset originating from an actual online-offline co-location deployment scenario. The other datasets leverage Azure LLM Inference Traces 2024~\cite{stojkovic2025dynamollm} for the online portion and combine them with offline requests from OOC dataset for the offline portion:
\begin{itemize}
    \item \textbf{OOC}: Our OOC dataset, includes real-world service traces of both online and offline requests.
	\item \textbf{Azure Conv}: AzureLLMInferenceTrace\_conv\_1week (2024) as the online request trace, with offline requests from OOC dataset.
	\item \textbf{Azure Code}: AzureLLMInferenceTrace\_code\_1week (2024) as the online request trace, with offline requests from OOC dataset.
\end{itemize}

The average input and output lengths of these datasets are summarized in Table~\ref{tab:avg_io}.

\begin{table}[h]
\small
\centering
\caption{Average prompt and output lengths across datasets.}
\label{tab:avg_io}
\begin{tabular}{lrr}
\toprule
Dataset & Avg. Prompt Length & Avg. Output Length \\
& (tokens) & (tokens) \\
\midrule
OOC (Online)   & 1892.47 & 1062.62 \\
OOC (Offline)  & 1200.52 &   671.51 \\
Azure Conv   & 1512.30 &    98.75 \\
Azure Code   & 2317.18 &    22.74 \\
\bottomrule
\end{tabular}
\end{table}
\subsubsection{Dataset Scaling}

For online requests, we use trace data with request timestamps to capture the real-world fluctuations in request arrival rates. However, since these datasets come from services of different scales, they exhibit varying overall request rates (or loads). If applied directly, they may either overload or underutilize our test environment, making it impossible to obtain meaningful SLO evaluation results.

To address this, we scale the request timestamps in the traces:
\begin{itemize}

	\item By randomly dropping requests at a fixed ratio, we reduce the overall request rate.
	\item By replicating existing request prompt and output lengths while interpolating their timestamps, we increase the overall request rate.
\end{itemize}

This scaling method only affects the aggregate request rate to fit the scale of our test cluster, without altering the temporal patterns of workload fluctuations. For example, a traffic spike lasting 5 minutes in the original trace will still last 5 minutes after scaling, and the ratio between peak and trough request rates is preserved. This ensures that, aside from service scale adjustments, our experiments remain consistent with the behavior of real-world services.

\subsubsection{Baseline} We use the standard xLLM as one of our baselines. Since, to the best of our knowledge, no existing open-source framework currently supports online–offline co-location under a disaggregated architecture, and prior work on co-location with non-disaggregated systems has not been open-sourced, we construct equivalent baselines on top of xLLM to match existing scheduling strategies. In the following experiments, the compared systems are defined as:
\begin{itemize}
    \item \textbf{base P/D}: Represents the P/D-disaggregated of standard xLLM framework, where both online and offline requests are treated as ordinary (online) requests. This is similar to applying existing P/D-disaggregated frameworks such as vLLM~\cite{kwon2023efficient}, SGLang~\cite{zheng2024sglang}, or DistServe~\cite{zhong2024distserve} directly in an online–offline co-location scenario.
	\item \textbf{online priority}: Represents an extension of base P/D with basic online/offline-aware scheduling. It only schedules offline requests when resources are idle, restricts the Decode batch size to safeguard online SLOs, and preempts offline requests during online traffic spikes. This setting is equivalent to migrating existing online-offline co-location systems (e.g., HyGen~\cite{sun2025hygen}, Echo~\cite{wang2025echo}) to a P/D-disaggregated framework.
	\item \textbf{\sysname{} (ours)}: Our proposed online–offline co-located inference system designed specifically for disaggregated architectures. It extends the original P/D disaggregation into latency-constraint disaggregation to maximize offline request utilization, and employs performance bottleneck–aware scheduling to ensure SLO compliance while maximizing the efficiency of compute, memory bandwidth, and memory capacity utilization.
\end{itemize}

\subsection{Online / Offline Co-location Serving Experiment}

In this section, we construct an online–offline service scenario and evaluate the systems described above to measure the maximum achievable offline request throughput without violating online request SLOs.

For each model–dataset combination, we first determine the traffic scaling factor for online requests such that the system can just meet the online traffic peak without SLO violations. This represents the resource utilization limit for a pure online service scenario. Although resource utilization may be under-saturated during off-peak periods, reducing resources would otherwise cause SLO violations at traffic spikes. Importantly, this also means that we do not allocate any additional resources for offline workloads.

Next, we gradually increase the offline workload traffic starting from zero. Since offline requests have no strict temporal constraints, we regulate their traffic using a uniform QPS-based control. For each level of offline load, we measure the impact on online request SLOs. If the SLO violation rate for online requests increases as offline load rises, it indicates that online processing has been interfered with by offline requests. Once the online request SLO violation rate exceeds the prescribed threshold (3\%), the system can no longer provide valid online services under that offline load. The offline throughput just before this point is defined as the maximum effective offline request throughput for the system.

\begin{figure*}[t]
  \centering
  \includegraphics[width=0.85\linewidth]{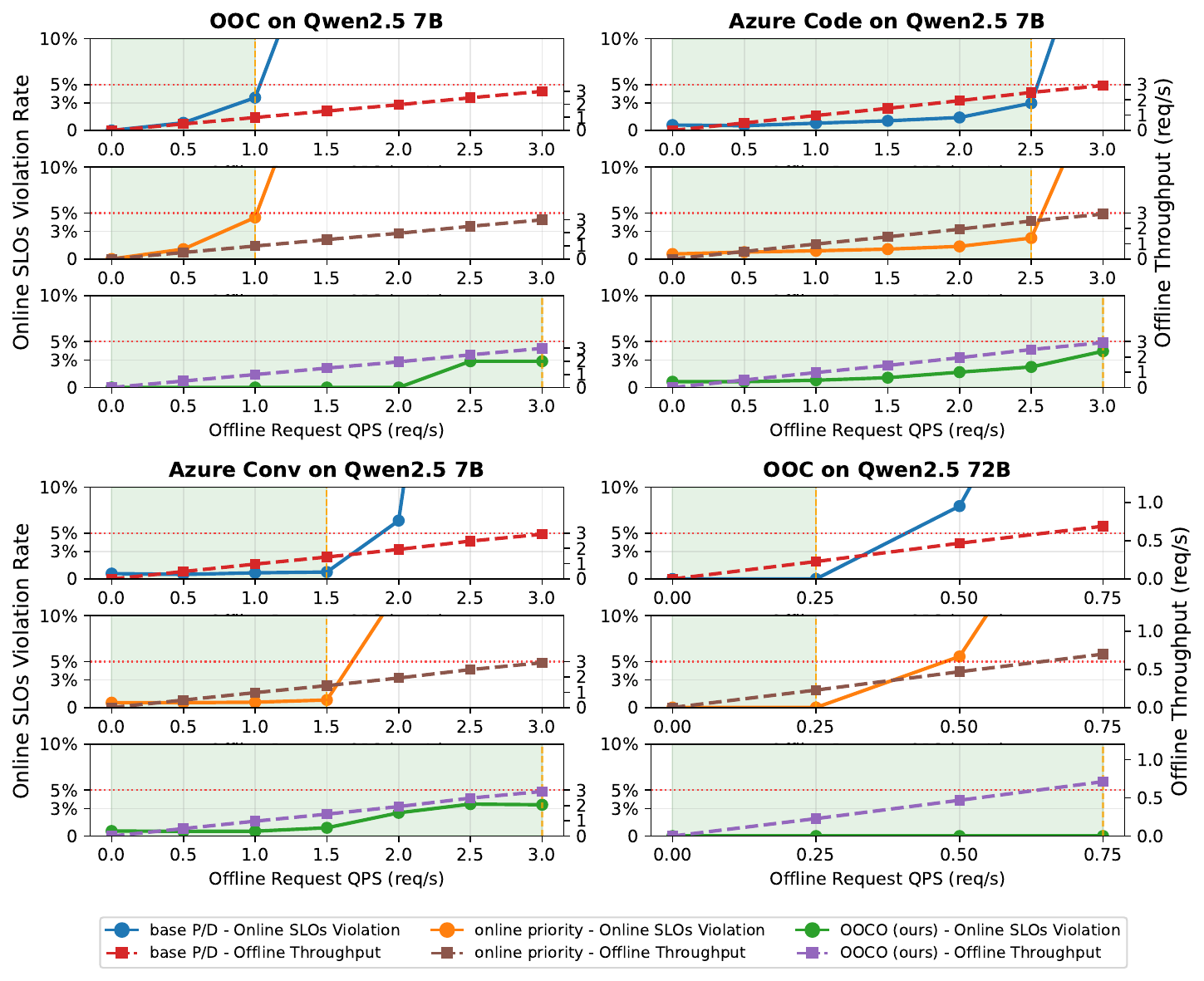}
  \caption{Online-offline co-location service experiment.}
  \label{fig:expr_main}
\end{figure*}

Our experimental results are shown in Figure~\ref{fig:expr_main}, where the green region indicates the range of offline request throughput that can be sustained without exceeding the violation threshold of online SLOs. As can be observed, both \textbf{base P/D} and \textbf{online priority} exhibit a sharp increase in online SLO violation rate once the offline request QPS surpasses a certain level, indicating that online requests are being interfered with by offline workloads. In contrast, \sysname{} maintains stable SLO performance even as offline request QPS continues to rise, achieving 1.17× to 3× improvement over the best-performing baseline across multiple datasets and models.

We also note that in our experiments, \textbf{online priority} did not outperform \textbf{base P/D}. We attribute this to its inability to efficiently utilize cluster resources compared with \sysname{}. Although \textbf{online priority} enforces protections for online requests, the inclusion of offline requests in decoding steps prolongs iteration latency. While this added latency does not immediately violate SLO constraints, it slows down the overall processing rate of online requests. Consequently, more requests remain resident in the system before online traffic peaks arrive, leading to queuing and degraded SLO performance during traffic surges.


\subsection{Baseline Evaluation}
he core contribution of \sysname{} lies in its system and scheduling design, which makes it portable to other hardware platforms and frameworks. Therefore, our experimental results based on NPU hardware and the xLLM framework are also of broader reference value. To further validate this point, we conducted a baseline evaluation to assess the effectiveness of our experimental platform.

Specifically, we used the same request set (from Azure Conv) to benchmark vLLM @ NVIDIA H800, vLLM @ Ascend 910c, and original xLLM @ Ascend 910c, all running Qwen2.5 7B. We measured throughput performance under single GPU/NPU deployment (single chip for 910c) in non-disaggregated mode, and pushed requests at maximum rate to stress the system to its throughput limit. This setup eliminates the influence of framework-level scheduling differences and isolates the performance of hardware and operator implementations. The results are shown in Table~\ref{tab:baseline_eval}.

\begin{table}[h]
\centering
\small
\caption{Maximum throughput comparison of Qwen2.5-7B across frameworks and hardware.}
\label{tab:baseline_eval}
\begin{tabular}{l r}
\toprule
Framework / Hardware & Throughput (token/s) \\
\midrule
vLLM @ NVIDIA H800  & 36099.72 \\
vLLM @ Ascend 910c (single chip) & 10050.44 \\
xLLM @ Ascend 910c (single chip) & 12083.43 \\
\bottomrule
\end{tabular}
\end{table}

We observe that vLLM on NVIDIA H800 achieves about 3× the throughput of vLLM on a single Ascend 910c chip, which is consistent with their theoretical peak FLOPs/s ratio. Moreover, xLLM on Ascend 910c outperforms vLLM on the same hardware, indicating that the combination of xLLM and 910c delivers inference performance at a mainstream level, with strong optimization of xLLM for the 910c platform. Therefore, adopting xLLM on Ascend 910c as our primary experimental platform is both reasonable and representative.

\section{Discussion and Related Work}
\label{sec:discussion}

With the rapid growth of LLM infrastructures, recent years have witnessed a surge of research efforts on optimizing LLM serving across diverse directions and technical routes. Some of these are closely related to our work, while others are orthogonal. We discuss them in detail below.

About the online-offline co-locating inference for LLMs, such as Echo~\cite{wang2025echo} and HyGen~\cite{sun2025hygen}. These systems are designed for non–disaggregated architectures, incorporating basic mechanisms such as online request prioritization, preemption, and SLO guarantees. However, if directly ported to a P/D-disaggregated system, they still fail to fully exploit cluster resources. In our experiments, we adopt such methods as one of our baselines, and demonstrate that our approach — restructuring the P/D disaggregation into a latency-constrained decomposition and applying performance bottleneck aware scheduling — is essential for P/D-disaggregated architectures. HyGen further introduces request reordering for offline jobs to better exploit prompt prefix cache under high prefix redundancy. While this technique is orthogonal to our focus on resource utilization efficiency, and hence not enabled in our experiments, it is worth noting that not all offline workloads exhibit high prefix redundancy (e.g., our OOC dataset). Moreover, our \sysname{} design can be extended to support request reordering when applicable.

About the traditional DNN co-locating inference, where systems such as Orion~\cite{strati2024orion} and Tally~\cite{zhao2025tally} address co-location of online and offline jobs or workloads with different priorities. However, their scheduling and inference characteristics differ significantly from LLMs. These works primarily focus on isolating resources across different models, whereas in LLM co-location, both online and offline requests share the same model. This fundamental difference makes their methods not directly applicable to our scenario.

About the dynamic adjustment of Prefill / Decode resource ratios in P/D-disaggregated inference frameworks, as explored in works such as Splitwise~\cite{patel2024splitwise} and Arrow~\cite{wu2025arrow}. These approaches handle fluctuations in P/D load balance for online requests by migrating jobs or dynamically switching node roles. Nevertheless, such methods face inherent limitations that make them less effective in offline co-location scenarios where P/D balance fluctuates even more frequently due to bursty online traffic:
\begin{itemize}
    \item They are difficult to apply in heterogeneous clusters with hardware or parallelism optimized separately for Prefill and Decode. Switching node roles in such environments can alter performance characteristics and potentially cause SLO violations.
	\item The switching latency between P/D roles is constrained by decode-phase KV Cache residency, requiring either migration or natural completion, which limits responsiveness.
\end{itemize}

In contrast, our \sysname{} framework tackles the additional P/D imbalance introduced by offline co-location through dynamic scheduling and migration of offline requests, achieving maximal resource utilization without altering the P/D node ratio. For the original P/D fluctuations arising solely from online requests, our method is orthogonal and can be complemented with these dynamic P/D ratio adjustment techniques as a secondary, longer-term balancing mechanism.

\section{Conclusion}
In this work, we present \sysname{}, a latency-constrained disaggregated inference system designed to efficiently co-locate online and offline workloads for large language models. By extending conventional P/D disaggregation into a latency-constrained disaggregation and introducing performance bottleneck aware scheduling based on roofline model, \sysname{} maximizes resource utilization while preserving strict online SLO guarantees. Through experiments on real-world traces and hardware platforms, we demonstrate that \sysname{} achieves substantially higher offline throughput compared to existing baselines, without compromising online service quality.

\bibliographystyle{ACM-Reference-Format}
\bibliography{references}

\end{document}